\documentclass[12pt,psfig,showpacs,letterpaper]{revtex4}
\usepackage{geometry}
\usepackage{graphicx}
\usepackage{natbib}
\usepackage{amsmath}
\usepackage{amssymb}
\begin{document}
\bibliographystyle{acm}

\title{Non-equilibrium two-fluid plasmas can generate magnetic fields and flows {simultaneously}}

\author{Hamid Saleem \\
National Centre for Physics (NCP),\\
Quaid-i-Azam University Campus,Islamabad, \\
Pakistan.}
\date{16, July 2010}

\begin{abstract}
A new analytical solution of the set of highly nonlinear two-fluid equations is
presented to explain the mechanism for the generation of "seed"
magnetic field and plasma flow by assuming the density n to have a
profile like an exponential in xy-plane and temperature profiles of
electrons (ions) to be linear in yz-plane. {It is shown that the
baroclinic vectors - $\nabla\Psi\times\nabla T_{j}$ (where $\Psi =
ln\overline{n}, \overline{n}$ is normalized density, and $T_{j}$
denote the temperatures of electrons and ions for j = e, i) can
generate not only the magnetic field but the plasma flow as well.}
It is also pointed out that the electron magnetohydrodynamics (EMHD)
model has inconsistencies because it does not take into account the
ion dynamics while the magnetic field is produced on slow time
scale. The estimate of the magnitude of the magnetic field in a
classical laser plasma using this model is in agreement with the
experimental observations.
\end{abstract}


\maketitle
\pagebreak

$\textbf{I. INTRODUCTION}$\\
The presence of large scale magnetic fields in galaxies, galaxy
clusters and in intergalactic space [1] is a mystery and several
theoretical models have been presented to explain the origin of
these fields [2-5]. Most of these works deal with the dynamo theory
of single fluid magnetohydrodynamics (MHD). But the set of MHD
equations assumes that some magnetic field is already present in the
system. Therefore, these models actually investigate the
amplification of the existing weak magnetic field and can not
explain the generation
of the 'seed' field in true sense. \\
Long ago [6], Biermann presented a mechanism for the generation of
stellar magnetic fields which is not based on MHD. He proposed that
the electrons faster motion compared to ions can produce electric
field in a rotating star which is not curl free due to non-parallel
density and temperature gradients and hence the magnetic field is
produced. The ions were assumed to be stationary in this work. The
Biermann battery and electron diffusion processes were also
investigated to
explain the generation of 'seed' magnetic fields in galaxies [7].\\
It is very interesting that the large magnetic fields of the order
of kilo and mega Gauss were observed in classical laser-induced
plasmas many decades ago [8-9]. These observations indicate that the
dynamics of initially unmagnetized nonuniform plasmas can generate
magnetic fields. The idea of magnetic field generation by plasma
dynamics is very attractive and a huge amount of research
work in this direction has already been appeared in literature.\\
Based on Biermann battery effect, a single fluid plasma model called
electron magnetohydrodynamics (EMHD) was presented to explain the
magnetic field generation in laser plasmas [10, 11]. But it does not
require the rotation of the system to produce magnetic fields. In
EMHD, ions are assumed to be stationary and electrons are treated to
be inertialess. Both the fluctuating [12, 13] and steadily growing
magnetic fields [8, 9] have been theoretically produced using EMHD
models. The so called magnetic electron drift (MEDV) mode was
discovered [12] using EMHD equations. The MEDV mode is believed to
be a pure transverse low frequency wave of an unmagnetized
inhomogeneous plasma. But a critical analysis of MEDV mode shows
that it should contain a contribution from electron density
perturbation as well.\\
Therefore, a new mode which is partially transverse and partially
longitudinal has been proposed to be a normal mode of pure electron
plasmas which can exist only in a very narrow range of parameters.
This mode can couple with the ion acoustic wave which also becomes
electromagnetic under certain conditions in a non-uniform plasma
[14]. Similarly the model equation widely used for the generation of
steadily growing magnetic field [12, 13] is not flaw-less. The field
grows on ion time scale while ions are assumed to be stationary.
Recently [15], the same model equation containing electron
baroclinic term has been used to estimate the magnetic
field produced in a laser plasma.\\
Some weaknesses and contradictions in the approximations and
assumptions used in EMHD models for magnetic field gereration have
already been pointed out [16]. The advantage of EMHD model is that
it is very simple. Since it is still being used by many authors,
therefore it seems important to discuss at least the two cases of
fluctuating and steadily growing magnetic fields which are believed
to be generated by EMHD. The EMHD models for the generation of
magnetic
fields are critically discussed in the next section.\\
The magnitudes of magnetic fields on galactic scale [7] as well as
on laser-plasma scale [8, 9] were estimated by assuming the
non-parallel electron density and temperature gradients to be
constant using EMHD equations. In these models the gradients of
electron temperature and density were assumed to be one-dimensional
and the produced magnetic fields had only one component.\\
A few years ago [17], a theoretical model was presented to show the
generation of three-dimensional magnetic field by baroclinic vectors
of electrons and ions. However, in this investigation some constant
magnetic field was assumed to be present already. A stationary
solution of these equations was presented by Mahajan and Yoshida
[18] in the form of double Beltrami field.\\
Later, the model presented in Ref. [17] was modified to explain the
creation of all the three components of 'seed' magnetic field vector
$\textbf{B}$ from t=0 due to externally given forms of baroclinic
vectors [19]. Here one does not need to assume some static magnetic
field to be present in the system. However, the
form of the solution was sinusoidal along one axis which is not physical in general.\\
The EMHD may have some applications in other areas but for the
magnetic field generation on slow time scale, the ion dynamics can
not be neglected. Therefore, it is necessary to study the 'seed'
magnetic field generation by using two fluid model. On the other
hand the generated magnetic field vector should not have necessarily
only one component for the sake of generality.\\
Our aim is to find out an analytical solution of the set of two
fluid equations such that the cross products of plasma density and
temperature gradients of electrons and ions become the source terms
in the electron and ion equations of motion. We assume that the
plasma has been produced in a non-equilibrium state and it evolves
with time generating the "seed" magnetic field and flow.\\
The present investigation is very different from the previous work
[19] because the density gradient scale length is assumed to have an
exponential form in xy-plane. In most of the analytical studies, the
density is assumed to have exponential form along one axis. But we
want to obtain a two-dimensional solution, therefore the density is
assumed to be a function of (x,y)
coordinates.\\
We have chosen special profiles of gradients of density $(\nabla n)$
electron temperature $(\nabla T_e)$ and ion temperature $(\nabla
T_i)$ to obtain an analytical solution. Different profiles of
density and temperatures can be considered but then the numerical
simulation will be needed. In our formulism all the nonlinear terms
vanish and ultimately we obtain the two linear equations where the
terms $\nabla \psi \times \nabla T_j$ (i = e, i) with $\psi = ln
\bar{n}$ (where $\bar{n}$ is normalized density) become the source
terms for
magnetic field.\\
The details of the model are discussed in section III. Since the
present model contains a very complex system of highly nonlinear
equations, therefore to find out an analytical solution one has to
use some assumptions and approximations. The main focus is to
justify the physical idea that the two fluid plasma with density
gradient like an exponential function in xy-plane and constant
gradients of electron and ion temperatures along y and z-axes can
generate the 'seed' magnetic fields and flow. This model can explain
the magnetic fields produced in laser-induced plasmas. In our
opinion, the numerical simulation of two-fluid equations is very
important to study the 'seed' field generation. In simulation one
can use many different profiles of density and temperatures.\\

$\textbf{II. CONTRADICTORY RESULTS OF EMHD}$\\
Here we briefly point out the contradictory results of EMHD models
used for magnetic field generation. First we discuss a theoretical
model based on EMHD for the generation of fluctuating magnetic
fields proposed several years ago [12]. The mode-discovered through
EMHD theory was named as the magnetic electron drift vortex (MEDV)
mode. A great deal of research
work on this mode has been carried out.\\
The critical analysis of MEDV mode and some new theoretical results
have been published recently on the fluctuating magnetic fields
[14]. There seems to be a need to briefly clarify here the physical
situation to lay down the basis of our theoretical model presented
in the next section. In the theory of MEDV mode, the ions are
assumed to be stationary but the electron inertial effects are
included. The linear description of the MEDV mode is presented very
briefly as follows.\\
Electron equation of motion is, $$m_e n_0 \partial_t
\textbf{v}_{e1}=-en_0 \textbf{E}_1 - \nabla p_{e1}\eqno{(1)}$$ where
subscripts one (1) and naught (0) denote the linearly perturbed and
equilibrium quantities, respectively. In the limit
$\omega_{pi}<<\omega << \omega_{pe}$ (where $\omega$ is the
frequency of the wave and $\omega_{pj}=\left(\frac{4\pi
n_{j0}e^2}{m_j}\right)^{1/2}$ is the plasma oscillation frequency of
the jth species while j= e here), the displacement current is
ignored and Maxwell's equation yields,
$$\nabla \times \textbf{B}_1=\frac{4\pi}{c}(\textbf{J}_1)=\frac{4\pi}{c}(-en_0
\textbf{v}_{e1})\eqno{(2)}$$ Since $\nabla.\textbf{J}_1=0$,
therefore according to (2), the density perturbation is neglected
and we find $p_{e1}=n_0 T_{e1}$. For $T_{e1}$, the electron energy
equation becomes,
$$\frac{3}{2}n_{0}\partial_{t}T_{e1}+\frac{3}{2}n_0
(\textbf{v}_{e1}.\nabla)T_{e0}=-p_{0}\nabla.\textbf{v}_{e1}\eqno(3)$$
Assuming, $\nabla n_0=\hat{\textbf{x}}\left|\frac{dn_0}{dx}\right|$,
$\kappa_n = \left|\frac{1}{n_0}\frac{dn_0}{dx}\right|$,
$\textbf{E}_1=E_1 \hat{\textbf{x}}$, $\textbf{k}=k_y
\hat{\textbf{y}}$ and $\textbf{B}_1=B_1 \hat{\textbf{z}}$ the linear
dispersion relation for MEDV mode turns out to be
$$\omega^{2}=\frac{2}{3}C_{0}(\frac{\kappa_{n}}{k_{y}})^{2}v^{2}_{Te}k_{y}^2\eqno(4)$$
where
$C_0=\frac{\lambda_{e}^{2}k_{y}^{2}}{1+\lambda_{e}^{2}k_{y}^{2}}$,
$\lambda_e=\frac{c}{\omega_{pe}}$ is the electron collision-less
skin depth and $\nu_{Te}=\left(\frac{T_e}{m_e}\right)^{1/2}$ is the
electron thermal speed. If temperature gradient is assumed to be
anti-parallel to the density gradient in laser plasma with $\nabla
T_0=\hat{\textbf{x}}\left|\frac{dT_0}{dx}\right|$ and
$\kappa_T=\left|\frac{1}{T_0}\frac{dT_0}{dx}\right|$, then (4) is
modified as,
$$\omega^2=C_0\frac{\kappa_n}{\kappa_y}\left[\frac{\left(\frac{2}{3}\kappa_n-\kappa_T\right)}
{k_y}\right]\nu_{Te}^{2}k_{y}^{2} \eqno{(5)}$$ and these magnetic
perturbations become unstable if the condition
$$\frac{2}{3}\kappa_n < \kappa_T\eqno{(6)}$$ holds. Note that the local approximation
requires $\kappa_n, \kappa_T << k_y$. It has been assumed that
$\nabla.\textbf{E}_1=0$ and $\nabla.\textbf{v}_{e1}\neq 0$ in
the description of MEDV mode along with $\omega_{pi}<<\omega$.\\
Equation (1) indicates that the term $\nabla p_{e1}=\nabla(n_0
T_{e1})=T_{e1}\nabla n_0+n_0 \nabla T_{e1}$ will produce a linear
term with $\nabla$ replaced by $\textbf{k}$ and hence $\textbf{E}_1$
can have a longitudinal component with $\nabla. \textbf{E}_1\neq 0$.
Thus the mode can not be a pure transverse
mode.\\
Moreover, the linear theory has been applied under the local
approximation therefore the term,
$\left(\frac{\kappa_n}{k_y}\right)^2 v_{T_e}^{2}k_{y}^{2}$ can be
closer to $c_{s}^{2}k_{y}^{2}$ where $c_s=(T_e/m_i)^{1/2}$ is the
ion acoustic speed while $C_0 < 1$ always holds. Using the laser
plasma parameters [9] $T_e=100 eV$ and $n_0 \sim 10^{20} cm^{-3}$,
one obtains $\omega<\omega_{pi}$ contrary to initial assumption of
stationary
ions for $\omega_{pi}<<\omega$.\\
Recently [14], it has been shown that if compressibility effects are
also taken into account then one obtains a new partially transverse
and partially longitudinal normal mode of a nonuniform pure electron
plasma as,
$$\omega^2=\frac{2}{3H_0}\lambda_{e}^{2}k_{y}^{2}(v_{te}^{2} \kappa_n^2)
\left(1-\frac{3}{2}\frac{\kappa_T}{\kappa_n}\right)\eqno{(7)}$$
where $H_0=\left[\left\{1+\frac{5}{3} \lambda_{De}^{2}
k_{y}^{2}\right\}a-\kappa_{n}^{2}/
k_{y}^{2}\right]$ and $a=(1+\lambda_{e}^{2}k_{y}^{2})$.\\
It is important to note that ions can be assumed to be stationary in
the limit $\frac{m_e}{m_i}\rightarrow 0$. Since for hydrogen plasma
$\frac{m_e}{m_i}\sim 10^{-3}$, therefore (7) is valid for $m_e/m_i <
\lambda_{De}^{2} k_{y}^{2}$, $\kappa_{n}^{2}/k_{y}^{2}$ and
$\omega^2 << \omega_{pe}^{2}$.\\
But the important point is that the term
$\nu_{Te}^{2}\kappa_{n}^{2}$ can be closer to $c_{s}^{2}k_{y}^{2}$
and lesser than $\omega_{pi}^{2}$ therefore the dynamics of ions
should not be ignored.\\
It is also interesting to mention here that ion acoustic wave (IAW)
has always been treated as a low frequency electrostatic mode. The
reason is that in the limit $\frac{m_e}{m_i}\rightarrow 0$, the
inertia-less electrons are assumed to follow the Boltzmann density
distribution in the electrostatic field $\textbf{E}=-\nabla \varphi$
as,
$$\frac{n_e}{n_0}\simeq e^{-e\varphi/T_e}\eqno{(8)}$$
In an inhomogeneous plasma we may have
$\frac{m_e}{m_i}<\kappa_{n}^{2}/k_{y}^{2}$ and in this case electron
inertia should not be neglected. Then for $\frac{m_e}{m_i}<
\lambda_{De}^{2} k_{y}^{2}$, the IAW follows the dispersion relation
[24],
$$\omega^2=c^{2}_{s}k_{y}^{2}\frac{(a-\kappa_{n}^{2}/k_{y}^{2})}{(ab-\kappa_{n}^{2}/k_{y}^{2})}\eqno{(9)}$$
where $b=\left(1+\lambda_{De}^{2}k_{y}^{2}\right)$. Hence inhomogeneous plasmas can have a low frequency
electromagnetic wave on ion time scale.\\
When the electron temperature perturbation effect is taken into
account, then modes described in (7) and (9) will couple to produce
a partially longitudinal and partially transverse wave with the
dispersion relation [24],
$$\omega^2=\frac{5}{3H_0}[(\lambda_{e}^{2}k_{y}^{2})\nu_{Te}^{2}\kappa_{n}^{2}\left(\frac{2}{3}-
\frac{\kappa_T}{\kappa_n}\right)+\left(a-\frac{\kappa_{n}^{2}}{k_{y}^{2}}\right)$$$$c_{s}^{2}k_{y}^{2}\left\{
\frac{5}{3}-\left(\frac{k_{T}^{2}}{k_{y}^{2}}+\frac{\kappa_T
\kappa_n}{k_{y}^{2}}\right)\right\}]\eqno{(10)}$$

If $\nabla p_{e0} = 0$ is used as the steady state condition the
above equation yields a basic low frequency electromagnetic wave of
inhomogeneous unmagnetized plasmas with the dispersion relation,
$$\omega^2 = \frac{5}{3H_0} \left[(\lambda_e^2 k_y^2) v_{te}^2
\kappa_n^2 + \left(a-\frac{\kappa_n^2}{k_y^2}\right) c_s^2
k_y^2\right]\eqno{(11)}$$ This wave has not been studied in plasmas
so far. In our opinion it can play very important role in the
generation of magnetic fluctuations in unmagnetized plasmas due to
several linear and nonlinear mechanisms. In a pure electron plasma
where ions are assumed to be stationary, (10) reduces to $$\omega^2
= \frac{5}{3H_0} (\lambda_e^2 k_y^2) v_{te}^{2}
\kappa_n^2\eqno{(12)}$$ But in our point of view the electron plasma
wave frequency in (12) is near ion acoustic wave frequency $c_s^2
k_y$ and hence it couples with it.\\

Now we look at EMHD theory for the generation of 'seed' magnetic
field which is steadily growing. Again ions are assumed to be
stationary in the time scale $\tau << \omega_{pi}^{-1}$. In addition
to this the electron inertia is also neglected assuming
$\omega_{pe}^{-1} << \tau$ . Then electron equation of motion
becomes, $$0 \simeq -e \textbf{E} - \frac{\nabla
p_e}{n}\eqno{(13)}$$ The faraday law is, $$\partial_t \textbf{B}= -c
\nabla \times \textbf{E}\eqno{(14)}$$ If it is assumed that $\nabla
n_0 =\hat{\textbf{x}} \left|\frac{dn_0}{dx}\right|$ and $\nabla T_0
= \hat{\textbf{y}} \left|\frac{dT_o}{dx}\right|$, then (13) and (14)
yield,
$$\partial_t \textbf{B} = -\frac{c}{e} \left(\frac{T_e}{L_n
L_T}\right)\hat{\textbf{z}}\eqno{(15)}$$
where $L_n = \kappa_{n}^{-1}$ and $L_T=\kappa_T^{-1}$ are constants.\\
Equation (15) is integrated from $\tau = 0$ to $\tau =
\frac{L_n}{c_s}$ to have [9 - 11], $$\textbf{B}=
\left\{\frac{c}{e}\left(\frac{T_e}{L_n L_T}\right)\tau\right\}
\hat{\textbf{z}}\eqno{(16)}$$ This is a well-known equation in
laser-plasma literature. Assuming $T_e \sim 100 eV$, $n_0 \sim
10^{20} cm^{-3}$, $L_n \sim L_T \sim 0.005 cm$, one obtains $c_s
\sim 3 \times 10^7 cm/Sec$ and hence $|B| \sim 0.6 \times 10^6$
Gauss [9]. Note that $\tau =\frac{L_n}{c_s} \simeq 1.66 \times
10^{-10}$ Sec and $\omega_{pi} \sim 1.3 \times 10^{13} rad/Sec$
while the laser pulse duration is of the order of a nano second.
Thus we have $\omega_{pi}^{-1} << \tau$ contrary to initial
assumption of stationary ions for $\tau << \omega_{pi}^{-1}$. An
equation similar to (16) has also been used to estimate the 'seed'
magnetic field generated by ionized clump of a galactic cloud [7].
The density gradient of the cloud has
been assumed to have exponential form.\\
The brief overview of EMHD models shows clearly that the theoretical
models for both the fluctuating and steadily growing magnetic fields
suffer from serious contradictions. Since the EMHD is still being
used [15] for estimating magnetic fields produced in laser plasmas,
therefore the weaknesses and contradictions have
been elaborated here again.\\
Since the plasmas have generally exponential density profiles,
therefore there is a need to find out an exact 2-D solution of the
set of two fluid equations assuming exponential type density
structure in a plane. It may also be mentioned here that in many
tokamak plasmas, the density falls almost exponentially near the
walls which gives
rise to drift waves. \\

$\textbf{III. EXACT SOLUTION OF 2-FLUID EQS.}$\\
Our aim is to search for an exact analytical solution of the set of
highly nonlinear partial differential equations of electron-ion
plasma to show how the system from a non-equilibrium state can
evolve in time generating the
'seed' magnetic field.\\

In our opinion the same physical mechanism is applicable at both
astrophysical and laboratory scales. Biermann [6] gave the
pioneering idea that the electron baroclinic vector $(\nabla n_e
\times \nabla T_e)$ can generate magnetic fields in rotating stars.
We just modify it a little by proposing that the 'seed' magnetic
field is a macroscopic phenomenon and it is generated on longer
spatial and temporal scales. Hence the ion dynamics can not be
neglected. Therefore, the ion baro-clinic vector $(\nabla n_i \times
\nabla T_i)$ is also crucial to be considered. However, the
quasi-neutrality approximation in slow time scale is a valid
approximation, therefore we use $n_i \sim n_e = n$. In thermal
equilibrium, the source of
magnetic field generation disappears.\\
For an analytical solution of a complex set of equations, we have to
use some assumptions and approximations. The exact solution
presented here contains exponential type density profile in xy-plane
which is the main deviation from the previous models [17, 19]. It is
important to note that the nonlinear terms do not vanish if we
assume exponential density fall or rise along both the axes x and y
as has been discussed in
section II.\\
Therefore, we have to choose a very special exponential function in
xy-plane for density which reduces the nonlinear equations into two
linear equations. The detailed mathematical model is presented
here.\\

The electrons are assumed to be inertialess in the limit
$|\partial_t|<<\omega_{pe}$, $c|\nabla|$ where
$\omega_{pe}=\left(\frac{4\pi n_0 e^2}{m_e}\right)^{\frac{1}{2}}$ is
the electron plasma frequency and c is the speed of light. We define
four scalar fields $\varphi,u,\chi$ and h such that the ion velocity
$\textbf{v}_i$ and magnetic field are defined, respectively, as [17,
19],
$$\textbf{v}_i=(\nabla\varphi\times\mathbf{\hat{z}}+u\mathbf{\hat{z}})f(t)=(\partial_y \varphi, -\partial_x \varphi,
u)f\eqno{(17)}$$
$$\textbf{B}=(\nabla\chi\times\mathbf{\hat{z}}+h\mathbf{\hat{z}})f(t)=(\partial_y
\chi, -\partial_x \chi, h)f\eqno{(18)}$$ All these scalar fields are
functions of (x, y) coordinates and f is a function of time.\\
We further assume $\partial_t n_j=0$ and $\nabla.\textbf{v}_j=0$
which requires $$\nabla\psi.\textbf{v}_j=\{\varphi,
\psi\}=0\eqno{(19)}$$ where $\psi=ln \bar{n}$, $n_e\simeq n_i =n$
and $\{\varphi, \psi\}=\partial_y \varphi \partial_x \psi-\partial_x
\varphi \partial_y \psi$. Here $\bar{n}=\frac{n_{(x,y)}}{N_{0}}$ and $N_{0}$ is an arbitrary number used to normalize n.\\
The displacement current is neglected and hence we obtain,
$$\textbf{v}_{e}=\left(\textbf{v}_{i}-\frac{c}{4\pi e}\frac{\nabla\times\textbf{B}}{n}\right)\eqno{(20)}$$ Let
$\textbf{E}=-\nabla\Phi-\frac{1}{c}\partial_t \textbf{A}$ where
$\Phi$ is electrostatic potential different from $\varphi$. Since
$\textbf{B}=0$ at t=0, therefore we do not normalize the equations.
This point has been explained in detail in Ref. [19]. The curls of
momentum equations of electrons and ions yield, respectively,
$$\partial_t
\textbf{B}+\nabla\times\left[\textbf{B}\times\left(\textbf{v}_i-\frac{c}{4\pi
e}\frac{\nabla\times\textbf{B}}{n}\right)\right]=-\frac{c}{e}(\nabla\psi\times
\nabla T_e)\eqno{(21)}$$ and
$$\partial_t(a\textbf{B}+\nabla\times\textbf{v}_i)-\nabla\times[a(\textbf{v}_i
\times \textbf{B})+\textbf{v}_i \times (\nabla\times
\textbf{v}_i)]$$$$=\frac{1}{m_i}(\nabla\psi\times T_i)\eqno{(22)}$$
where $a=\frac{e}{m_i c}$. If the conditions
$$\{\varphi,\chi\}= \{\varphi, u\}= \{h, \varphi\}=0\eqno{(23)}$$ satisfy along with $$\{\nabla^2 \varphi,
\varphi\}=0\eqno{(24)}$$ then all the nonlinear terms of (21) and
(22) vanish and they reduce, respectively, to simpler equations
$$\partial_t \textbf{B}=-\frac{c}{e}(\nabla \psi\times \nabla
T_e)\eqno{(25)}$$ and $$\partial_t
(a\textbf{B}+\nabla\times\textbf{v}_i)=\frac{1}{m_i}(\nabla\psi\times\nabla
T_i)\eqno{(26)}$$ where $T_e\neq T_i$ and right hand sides of (25)
and (26) are the source terms for generating magnetic field and
plasma flow. Let us assume that $\textbf{B}$ is related with plasma
vorticity through the following equation,
$$\textbf{B}=\alpha\left(\nabla\times\textbf{v}_{i}\right)\eqno(27)$$ where
$\alpha$ is a constant. Then (26) becomes
$$\left(a+\alpha^{-1}\right)\partial_{t}\textbf{B}=\frac{1}{m_{i}}\left(\nabla\psi\times\nabla
T_{i}\right)\eqno(28)$$ Now we discuss an important point of the
present theoretical model. In the previous works it was assumed that
the field $\varphi$ satisfies the Poisson equation,
$$\nabla^{2}\varphi=-\lambda\varphi\eqno(29)$$ where $\lambda$ is a
constant and $0<\lambda$ holds. The forms of $\psi$ and $T_{j}$ were
chosen as $\psi=\psi_{0}e^{\mu_{1}x}cos \mu_{2}y$ and
$T_{j}=\{T_{00j}+T^{'}_{0j}(y-z)\}f(t)$ where $\psi_{0}$, $\mu_{1}$,
$\mu_{2}$, $T_{00j}$ and $T^{'}_{0j}$ were constants. We, here, want
to find out a 2-D solution in the exponential form without assuming
the density gradient to be constant. For this purpose the assumption
(29) is modified as
$$\nabla^{2}\varphi=\lambda\varphi\eqno(30)$$ and we assume $0<\lambda$. The form
of $\psi_{(x,y)}$ can be chosen like, $$\psi_{(x,y)}=A_{1}e^{(\mu
x+\nu y)}+A_{2}e^{(\mu x-\nu y)}=\psi_{1}+\psi_{2} = ln
\bar{n}\eqno(31)$$ where $\overline{n}=\frac{n_{(x,y)}}{N_{0}}$ is
dimensionless and $N_{0}$ is some constant density. Here $A_{1}$,
$\mu$, $\nu$, $A_{2}$ are constants and
$$\lambda=\mu^{2}+\nu^{2}\eqno(32)$$ We may choose $0<\mu, \nu$ for
simplicity. Let the temperatures be only functions of space in
yz-plane as,
$$T_{0j}(y,z)=T_{00j}+T^{'}_{0j}(y-z)\eqno(33)$$ Then the baroclinic vectors in (25) and (26)
become constant with respect to time. These equations can be
integrated from t=0 to $\tau$ and one obtains,
$$\textbf{B}=-\frac{c}{e}(\nabla \psi \times \nabla
T_e)\tau\eqno{(34)}$$ and $$\textbf{B}=\frac{1}{m_i
(a+\alpha^{-1})}(\nabla \psi \times \nabla T_i)\tau\eqno{(35)}$$
These equations relate $T^{'}_{e0}$ and $T^{'}_{i0}$ as,
$$T^{'}_{e0}=\frac{a}{(a+\alpha^{-1})}T^{'}_{i0}\eqno{(36)}$$ Equations (31) and (33) yield
$$\left(\nabla\psi\times\nabla
T_{j}\right)=-T^{'}_{0j}\left(\partial_{y}\psi, -\partial_{x}\psi,
-\partial_{x}\psi\right)\eqno(37)$$ where
$\partial_{y}\psi=\nu\left(\psi_{1}-\psi_{2}\right)$ and
$\partial_{x}\psi=\mu\psi$.\\
We are free to use any of the equations, (34) or (35) to estimate
$\textbf{B}$. Let us choose (34) and use (18) for f=1 to find out
following relations,
$$\chi=(\chi_{0}\psi)\tau\eqno(38)$$ and $$h=-(h_{0}\psi)\tau\eqno(39)$$
where $\chi_{0}=\left(\frac{cT_{0e}^{'}}{e}\right)$
and $h_{0}=\mu\chi_{0}$. Equation (17) for f=1 along with (30)
gives,
$$\nabla\times\textbf{v}_{i}=\left(\partial_{y}u, -\partial_{x}u, -\lambda\varphi\right)\eqno(40)$$
Then (27) yields, $$u=(u_{0}\psi)\tau\eqno(41)$$ and
$$\varphi=(\varphi_{0}\psi)\tau\eqno(42)$$ where $u_{0}=\frac{\chi_{0}}{\alpha}$ and
$\varphi_{0}=\left(\frac{\mu}{\alpha\lambda}\right)\chi_{0}$.\\
Three dimensional magnetic field $\textbf{B}$ and $\textbf{v}_{i}$
can be
expressed explicitly for $A_1 \neq A_2$ as,\\

\[\textbf{B}= \left[ \begin{array}{ c } \chi_{0}\nu
\left(A_{1}e^{(\mu x+\nu y)}+A_{2}e^{(\mu x-\nu y)}\right)  \\
-\mu\chi_{0}\left(A_{1}e^{(\mu x+\nu y)}+A_{2}e^{(\mu x-\nu
y)}\right) \\ -h_{0} \left(A_{1}e^{(\mu x+\nu y)}+A_{2}e^{(\mu x-\nu
y)}\right) \end{array} \right]\psi_0\]$$\eqno(43)$$ and
\[\textbf{v}_{i}= \left[ \begin{array}{ c }\nu \varphi_{0}(A_{1}e^{(\mu x + \nu y)}-A_{2}e^{(\mu x-\nu
y)})\\
-\mu \varphi_{0}(A_{1}e^{(\mu x + \nu y)}+A_{2}e^{(\mu x-\nu
y)})\\-u_{0}(A_{1}e^{(\mu x + \nu y)}+A_{2}e^{(\mu x-\nu
y)})\end{array}\right]\psi_0 \]$$\eqno(44)$$\\
Hence all the scalar fields $\chi, h, u$ and $\varphi$ become
functions of (x,y) through $\psi$ which is externally given. The
complicated nonlinear terms of two-fluid equations vanish and we
obtain two simple and beautiful linear equations (34) and (35).
These equations show that the terms of electrons and ions $(\nabla
\psi \times \nabla T_j)$ (for j = e, i)
become the source for the 'seed' magnetic field and plasma flow $\textbf{v}_i$.\\

{$\textbf{IV. GENERAL APPLICATIONS}$\\
It has been shown that the forms of $\psi (x,y)$ and $T_{0j} (y,z)$
given in equations (31) and (33), respectively, reduce the set of
nonlinear two fluid partial differential equations into two simpler
linear equations (34) and (35) under certain conditions mentioned in
the previous section. This theoretical model shows that the
non-parallel density and temperature gradients can create magnetic
fields and flows in initially unmagnetized
plasmas.\\
Our aim is to apply this model to a system which has
exponential-type of density profile as for example in the case of a
laser plasma. But the chosen form of $\psi (x,y)$ in (31) with the
definition $\psi = ln \bar{n}$ gives,
$$\bar{n}=\frac{n_{(x,y)}}{N_0}=exp\left[A_0 \left\{ e^{(\mu x + \nu y)} +
e^{(\mu x - \nu y)}\right\}\right] \eqno{(45)}$$ where $A_0 = A_1 =
A_2$ has been assumed.\\
It looks as if density $n_{(x,y)}$ has a profile like double
exponential in (x,y) plane. Such a steep density variation is not
interesting for physical applications, in general. We show here that
the density variation can become very similar to exponential form in
(x,y) plane by choosing suitable values of the constants $N_0$,
$\mu$ and $\nu$ along with $A_0$. Let us consider an inhomogeneous
plasma rectangle in (x,y) plane with four corner points (0,0),
$(x_m, 0), (0, y_m)$ and $(x_m,y_m)$ where $x_{m}$ and $y_{m}$ are
the maximum lengths of the system along x and y axis, respectively.
Then choose the constants in such a way that the density $n_{(x,y)}$
at $(x_m,y_m)$ will be almost e-times (or a little larger) than the
value at (0,0), while the density at
$(x_m,0)$ and $(0,y_m)$ will be somewhat lesser than e-times the density at (0,0).\\
Such a density function is acceptable physically. For example, in
previous EMHD model, the density was chosen to be an exponential
function only along x-axis as $n_{(x)}=n_0 e^{\frac{x}{L_n}}$ where
$L_n$ is the density scale length and $n_0$ is the magnitude of
density at x=0 [9]. In our case, $n$ depends upon two coordinates x
and y. It's profile depends upon the values of the constants.
Therefore this theoretical model can be applicable to many
inhomogeneous plasma systems.\\
Note that $\psi= ln \bar{n}$ and if $0<A_0 < 1$, $0\leq\mu x + \nu y
< 1$ and $0\leq\mu x - \nu y < 1$, then $\psi$  can have values
which give $e^{\psi}$ of the order of $e^{(1)}$, but not exactly
$e(1)
\simeq 2.7$ at all points because $\psi$ changes with x and y.\\
In the next section we shall apply the model to laser plasmas as an
example and our point of view will become clearer. The values of the
constants will be chosen to show how it works for relatively smooth
density profiles in (x,y)-plane.\\
It seems important to point out that any one of $\psi_1$ and
$\psi_2$ in (31) should not be much smaller than the other while
choosing the constants. If one of them is negligibly small then the
solution becomes one-dimensional.}

{$\textbf{V. LASER PLASMA}$\\
Here we apply our theoretical model to estimate magnetic field
$\textbf{B}$ and the plasma flow $\textbf{v}_{i}$ in a non-uniform
classical laser plasma. Consider a finite plasma rectangle with four
corner points $(0, 0)$, $(x_{m}, 0)$, $(0, y_{m})$ and $(x_{m},
y_{m})$ in (x, y) plane as has been mentioned in previous section.
We may assume $\mu x$ and $\nu y$ to vary in this finite plasma as,
$$\mu x: 0\rightarrow 0.5 = \mu x_{m}\eqno(46a)$$ and $$\nu y :
0\rightarrow 0.7 = y_{m}\eqno(46b)$$ Then at (0, 0), we have,
$\overline{n}_{(0, 0)}=\frac{n_{(0, 0)}}{N_{0}}$ and $N_{0}$ is
chosen such that $\overline{n}\neq1$ or $\overline{n} \nless 1$
because density $n_{(x, y)}$ should be neither zero nor negative.
Therefore, we choose $\frac{n_{(0, 0)}}{N_{0}}=3$ which gives,
$\Psi_{(0, 0)}\simeq1.1$ and due to (45) we find $A_{0}\simeq0.55$.
If density is of the order of $10^{20}cm^{-3}$, then we may assume
$N_{0}=10^{20}cm^{-3}$ and hence $n_{(0, 0)} =
3\times10^{20}cm^{-3}$. Or we may assume $N_{0}=10^{19}cm^{-3}$ and
hence $n_{(0, 0)}=3\times10^{19}cm^{-3}$ while $n_{(x_{m}, y_{m})}$
will turn out to be nearly $10^{20}cm^{-3}$. In laser-plasmas,
$L_{n}=50\times10^{-6}$ $m=L_{T}$ was assumed in estimating $|B|$
[8, 9] (where $L_{n}$ and $L_{T}$ are scale lengths of density and
electron temperature along x and y co-ordinates, respectively). In
these studies, the $\nabla n$ was along x-axis and $\nabla T_{e}$
was along y-axis only. Then $L_{n}\simeq L_{T}$ was also assumed.
For the sake of generality we do not assume $\mu = \nu$. Instead let
$\nu = 1.5 \mu$. In this case we estimate $\overline{n}$, $\Psi$ and
$\textbf{B}$ at 4-corner points of the plasma rectangle as follows:
$$\overline{n}_{(0, 0)}=3,\Psi_{(0, 0)}\simeq1.1$$
$$\textbf{B}_{(0, 0)}=(0, -1.1, -1.1)9.9\times10^{5} Gauss\eqno(47a)$$
$$n_{(x_{m}, 0)}=6:\Psi_{(x_{m}, 0)}\simeq1.79$$
$$\textbf{B}_{(x_{m}, 0)}=(0, -1.79, -1.79)9.9\times 10^{5}Gauss\eqno(47b)$$
$$\overline{n}_{(0, y_{m})}=3.98;\Psi_{(y_{m}, 0)}=1.38$$
$$\textbf{B}_{(0, y_{m})}=(1.24, -1.38, -1.38)9.9\times10^{5}Gauss\eqno(47c)$$
$$n_{(x_{m}, y_{m})}\simeq9.48;\Psi_{(x_{m},
y_{m})}\simeq2.25$$ $$\textbf{B}_{(x_{m}, y_{m})}\simeq(2.73, -2.25,
-2.25)9.9\times10^{5}Gauss\eqno(47d)$$ These values of
$|\textbf{B}|$ are almost in agreement with the observation [9].Then
we can express,
$$\textbf{v}_{i(x, y)}=6\times10^{7}(-0.46, 0.3, -1)\Psi_{(x, y)}cm/sec\eqno(48)$$
If we look at the values of density, we notice that density $n_{(x,
y)}$ at $(x_{m}, y_{m})$ is $$n_{(x_{m}, y_{m})}\simeq \{n_{(0,
0)}\}(3.16)$$ which is little larger than e-times the density at (0,
0). Therefore, the density has not been chosen as a double
exponential function. We are mainly interested in the order of
magnitudes of $|\textbf{B}|$ and $|\textbf{v}_{i}|$ which mainly
depends upon our choice of constants.\\}

$\textbf{V. DISCUSSION}$\\
It is important to note that the 'seed' magnetic field generation
can not be explained on the basis of magnetohydrodynamics (MHD).
There must be a source like thermal energy which converts into
magnetic energy. Biermann [6] proposed that the electron baroclinic
vector $(\nabla n_e \times \nabla T_e)$ can generate the magnetic
fields in rotating stars. Then based on this idea, a very simple
model; the electron magnetohydrodynamics (EMHD) was presented to
explain the generation of magnetic field in laser-induced plasmas.
Later on, the magnetic electron drift vortex (MEDV) mode was
discovered using EMHD. This was believed to be a low frequency pure
transverse normal mode of electron plasma [12, 13]. The EMHD is also
not a convincing theoretical model for the generation of 'seed'
magnetic field. The MEDV mode description also
suffers from contradictions.\\
On the other hand, the two-fluid model is too complicated. The
numerical simulation of these equations is a complex problem. But
the analytical 3-D solution is also not straightforward. However, a
two-dimensional solution has been presented using physical and
consistent assumptions and approximations. {It is important to note
that the present model is different from the previous works because
it shown that \begin{enumerate}
\item ion dynamics play a crucial role
\item the baroclinic vectors generate not only the magnetic field
but plasma flow as well
\end{enumerate}}
An exact 2-D solution of the two-fluid equations was also found a
few years ago [19], but it had a serious weakness. The density
gradient was assumed to follow sinusoidal behavior contrary to
common observations. Since, it was the first effort to get an exact
solution of the two fluid equations, therefore it was presented for
the interest of
researchers working in the field.\\
The present 2-D exact solution of two-fluid equations is in the form
of exponential function of x and y coordinates. This structure of
density gradient is more physical. Since all fields become linear
function of $\psi$, therefore all fields have the similar spatial
structure. This solution is applicable to both astrophysical and
laser-induced plasmas in our opinion.\\
The present investigation suggests that instead of EMHD, the
numerical simulation of two-fluid equations will be very useful for
understanding the mechanism for the generation of 'seed' magnetic
field in different systems with different profiles of density and
temperatures. For analytical solution, we have to choose very
special forms of density and temperature gradients. This theoretical
model can be very useful for further studies in astrophysical and
laser plasmas.

$\textbf{Acknowledgement}$\\
The author is grateful to Professor Zensho Yoshida of Tokyo
University for several useful discussions on this work at Abdus
Salam-International Centre for Theoretical Physics (AS-ICTP),
Trieste, Italy during the Summer College on Plasma Physics 10-28
August 2009.

\end{document}